\documentclass[aps,pra,twocolumn,superscriptaddress,longbibliography,nofootinbib]{revtex4-1}
\usepackage[normalem]{ulem}
\usepackage{float}
\usepackage{graphicx}  
\usepackage{dcolumn}          
\usepackage{amssymb}
\usepackage{quantikz}
\usepackage{appendix}
\usepackage{physics}   
\usepackage{dsfont}
\usepackage{mathtools}
\usepackage{esvect}
\usepackage{wrapfig}
\usepackage{amsthm}
\usepackage{verbatim}
\usepackage{bbm}
\usepackage{placeins}
\usepackage[colorlinks=true,linkcolor=blue,citecolor=red,plainpages=false,pdfpagelabels]{hyperref}
\usepackage{bm}
\usepackage{makecell}
\usepackage{multirow}
\usepackage{caption}
\usepackage{subcaption}

\usepackage[mathscr]{euscript}
\usepackage[normalem]{ulem}

\def\Tr{\operatorname{Tr}}

\def\({\left(}
\def\){\right)}
\def\[{\left[}
\def\]{\right]}

\newtheorem{theorem}{Theorem}

\newtheorem{definition}{Definition}

\newtheorem{observation}{Observation}

\newtheorem{note}{Note}

\def\>{\rangle}
\def\<{\langle}

\begin{document}

\title{A quantum information theoretic analysis of \\reinforcement learning-assisted quantum architecture search}

\author{Abhishek Sadhu}
\email{abhisheks@rri.res.in}
\affiliation{Raman Research Institute, Bengaluru, India}
\affiliation{Centre for Quantum Science and Technology (CQST), International Institute of Information Technology, Hyderabad, Telangana, India}

\author{Aritra Sarkar}
\email{a.sarkar-3@tudelft.nl}
\affiliation{QuTech, Delft University of Technology, Delft, The Netherlands}

\author{Akash Kundu}
\email{akash.kundu@helsinki.fi, corresponding author}
\affiliation{Joint Doctoral School, Silesian University of Technology, Gliwice, Poland}
\affiliation{Institute of Theoretical and Applied Informatics, Polish Academy of Sciences, Gliwice, Poland}
\affiliation{QTF Centre of Excellence, Department of Physics, University of Helsinki, Finland}

\begin{abstract}
In the field of quantum computing, variational quantum algorithms (VQAs) represent a pivotal category of quantum solutions across a broad spectrum of applications. These algorithms demonstrate significant potential for realising quantum computational advantage. A fundamental aspect of VQAs involves formulating expressive and efficient quantum circuits (namely ansatz), and automating the search of such ansatz is known as quantum architecture search (QAS). Recently, reinforcement learning (RL) techniques is utilized to automate the search for ansatzes, known as RL-QAS. This study investigates RL-QAS for crafting ansatz tailored to the variational quantum state diagonalisation problem. Our investigation includes a comprehensive analysis of various dimensions, such as the entanglement thresholds of the resultant states, the impact of initial conditions on the performance of RL-agent, the phase transition behaviour of correlation in concurrence bounds, and the discrete contributions of qubits in deducing eigenvalues through conditional entropy metrics. We leverage these insights to devise an entanglement-guided admissible ansatz in QAS to diagonalise random quantum states using optimal resources. Furthermore, the methodologies presented herein offer a generalised framework for constructing reward functions within RL-QAS applicable to variational quantum algorithms.
\end{abstract}

\maketitle

\section{Introduction}

Quantum computing represents a paradigm-shifting approach to computation, leveraging the principles of quantum mechanics to process information in ways fundamentally different from classical computers. 
At its core, this technology employs quantum bits, or qubits, which can exist in superposition states, enabling parallel computation paths. 
The information in qubits is transformed via quantum circuits consisting of quantum logic gates that are capable of acting on these computation paths in tandem.
The quantum circuits represent a corresponding quantum algorithm that can effectuate a computational advantage for a target application~\cite{harrow2017quantum,montanaro2016quantum,sadhu2023practical}.
These algorithms are either constructed manually, based on conventional logical reasoning or automatically by composing the quantum gates to replicate a required input-output behaviour via training.
The latter approach is termed quantum architecture search~(QAS)~\cite{zhang2022differentiable, du2022quantum, kuo2021quantum, zhang2021neural}.

QAS typically consists of two parts.
First, a template of the circuits is built, called the ansatz.
The ansatz can have parametric quantum gates, e.g., rotation angles.
Then, these parameters are determined via the variational principle using a classical optimiser in a feedback loop.
Algorithms constructed via this technique are called variational quantum algorithms~(VQA)~\cite{mcclean2016theory,cerezo2021variational}.
QAS can also be used to determine non-parametric circuits as an approach for quantum program synthesis~\cite{sarkar2024automated}.
In this research, however, we will focus on QAS in the context of VQA.

The ansatz design is critical, as it directly influences the expressivity and efficiency of the quantum solution.
Given the vast potential configuration space of quantum circuits, the process of identifying a suitable ansatz is a challenging problem. 
In~\cite{ostaszewski2021reinforcement, kundu2024enhancing, patel2024curriculum, du2022quantum}, the search for the quantum circuits that solve various VQA has been automated using reinforcement learning (RL) techniques~\cite{sutton2018reinforcement}. 
We particularly call ansatzes capable of solving a VQA an \textit{admissible ansatz}. 
The automated search for such an admissible ansatz using the RL-agent is termed reinforcement learning-assisted quantum architecture search (RL-QAS).
In this setting, after a certain number of episodes, the RL agent returns possible configurations of quantum ansatz that solve the problem. 
For $E_{tot}$ episodes, the agent returns $E_s$ successful episodes, and in $E_{tot}-E_s$ episodes, the agent fails to find a configuration within a specific predefined accuracy and depth. 
It has been shown in~\cite{ostaszewski2021reinforcement, wu2023quantumdarts,kundu2024enhancing,patel2024curriculum,kundu2024reinforcement} that classical machine learning-driven QAS algorithms outperform the state-of-art structures of ansatz and effectively optimise the number of parameters, depth, and impact of noise in VQAs.

The use-cases of such admissible ansatzes cover applications in quantum information theory~\cite{larose2019variational,cerezo2021variational,tan2021variational,cerezo2020variational, kundu2022variational}, chemistry~\cite{kandala2017hardware,mustafa2022variational,delgado2021variational} and combinatorial optimization~\cite{khairy2020learning,liu2022layer,glos2022space}.
Yet, a thorough investigation of such admissible ansatzes is still missing in recent literature. 
It has been previously shown~\cite{patel2024curriculum} that in RL-QAS methods, it is crucial to feed proper information about the problem to the agent to enhance the agent's performance. 
One of the ways of providing this information is by engineering a reward function. 
In RL, the formulation of the reward function and its engineering can be sparse or dense. The choice of optimal reward function depends on trial and error by investigating the performance of the agent under various reward configurations.

In this article, we numerically investigate the admissible ansatzes proposed by the RL-agent from the perspective of quantum information theory~\cite{wilde2013quantum}. For the investigation, we specifically focus on the reinforcement learning enhanced variational quantum state diagonalisation (RL-VQSD) problem~\cite{kundu2024enhancing}. In the VQSD algorithm~\cite{larose2019variational}, the problem-inspired ansatz is based on density matrix exponentiation~\cite{lloyd2014quantum}, but its depth and number of gates increase exponentially with the number of qubits. Hence, searching for an optimal ansatz (with a minimum number of gates and depth) that diagonalises a quantum state is an open research problem. In RL-VQSD, the search for the optimal configuration of ansatz is done by using a double deep-Q network (DDQN). See Section~\ref{appndx:rl-vqsd} for further details on the VQSD algorithm, the agent and environment specifications. The RL-agent proposes $E_s < E_{tot}$ admissible ansatzes that solves the VQSD problem. \textcolor{black}{The primary motivation of our work is to filter potential admissible ansatzes based on concurrence, using the highest and lowest entanglement bounds. This helps identify the range of optimal ansatzes and determine if increasing entanglement can improve the agent's performance. Another goal is to see if we can find an admissible ansatz that both diagonalizes and enhances the entanglement of the vacuum state}. 

Our investigation shows that the concurrence of the $E_s$ ansatz lies between an upper and lower bound, independent of the different initialisation of the double deep-Q network (DDQN), where each initialisation of DDQN corresponds to a different random initialisation of the weights. 
Moreover, through the numerical analysis we show that the upper and lower bounds are anti-correlated with respect to the increment in the initial state entanglement. 
The anti-correlation turns into a mild, and later strong, correlation when the initial state entanglement goes beyond $0.322$, indicating a phase transition in correlation between the upper and lower bounds of the concurrence. 
The phase transition provides us with an in-depth insight into the relation between the input and the output of the VQSD algorithm. 
Further, our investigation reveals that the RL-agent can generate high concurrence {admissible ansatzes} with fewer 2-qubit gates and circuit depth compared to ansatzes with very low concurrence. 
This is used as a guiding principle to design a two-part ansatz, where an entanglement enhancing initial block adjusts the input state's concurrence thereby enhancing the performance of the RL-agent by a factor of $2$.
Furthermore, we quantify the contribution of individual qubits of the RL-ansatz using conditional quantum entropy. Using such a measure, we observe a correlation between entanglement and quantum entropy, revealing why RL-VQSD can efficiently find the largest eigenvalues but fails to find the smallest ones.

The structure of the paper is as follows. \textcolor{black}{We present an overview of our main results in Sec.~\ref{sec:contribution}. 
In Sec.~\ref{appndx:rl-vqsd}, we present an overview of the reinforcement learning enhanced variational quantum state diagonalisation algorithm. We present an analysis of the sampled 2-qubit Haar-random quantum states in Sec.~\ref{appndix:input-state-analysis}.} We present the main results in Sec.~\ref{sec:mainResults}. Specifically, in Sec.~\ref{sec:entanglement}, we analyse the upper and lower bounds on the concurrence of the admissible ansatzes that solve the VQSD problem. In Sec.~\ref{sec:DQNentBound} we present the variation of the entanglement bounds for different weight initialisation of the DDQN. We observe the correlation properties of the entanglement bounds dependent on the concurrence of the initial state in Sec.~\ref{sec:correlationEntBound}. In Sec.~\ref{sec:agentEnvInteract}, we present the dependence of the entanglement bounds on the number of gates and circuit depth of the ansatz. The analysis of the contribution of individual qubits to the admissible ansatz is provided in Sec.~\ref{sec:conditionalEntropy}. In particular, we present the contribution of each qubit for change in the entanglement of the ansatz in Sec.~\ref{sec:entropyEntanglement}. In Sec.~\ref{sec:problem}, we explain why RL-VQSD can efficiently find the largest eigenvalues of the initial state but not the smallest eigenvalues. We provide concluding remarks and discuss possible future directions in Sec.~\ref{sec:discussion}.

\section{Contributions} \label{sec:contribution}
Through the following points, we summarise the contribution of the paper.
\begin{enumerate}
    \item \textcolor{black}{We numerically} demonstrate that the entanglement of the states generated by the RL-agent-proposed admissible ansatzes lies within a lower and upper concurrence bound. The upper and lower bounds of concurrence remain consistent across various weight initialisation of the DDQN. Furthermore, we show that these bounds are anti-correlated with respect to increasing entanglement until concurrence surpasses $0.322$; after this point, the anti-correlation turns into correlation. Hence, we introduce the concurrence $0.322$ as the phase transition point.
    \item Investigation of ansatzes in upper and lower bounds reveals that the optimal ansatz for VQSD generates strongly entangled states. The RL-agent requires fewer gates and circuit depth to generate such states to produce strongly entangled admissible ansatzes than weakly entangled ansatzes. We then utilise this information to minimise RL-agent and environment (which is the configuration of ansatz) interaction, thereby optimising the convergence time of the RL-VQSD. Specifically, we propose a two-part ansatz, where an entanglement enhancing initial block adjusts the input state's concurrence thereby enhancing the performance of the RL-agent by a factor of $2$.
    
    \item We \textcolor{black}{numerically} evaluate individual qubits' contribution in diagonalising 2-qubit quantum states using conditional quantum entropy. Our investigation shows that the qubits have equal contributions on average in diagonalising random quantum states.
    \item We further investigate the correlation in conditional entropy between qubits for different inferred eigenvalues.
    It requires mild \textcolor{black}{correlation} for the first two largest eigenvalues and \textcolor{black}{mild anti-correlation} for the smallest eigenvalues. Most admissible ansatzes lie in a mild \textcolor{black}{correlation} regime, explaining VQSD's limitation in finding the smallest eigenvalues.
\end{enumerate}

\section{The reinforcement learning enhanced variational quantum state diagonalisation}\label{appndx:rl-vqsd}

The VQSD algorithm~\cite{larose2019variational} for a quantum state $\rho$ comprises three subroutines: (1) In the training subroutine, the parameters $\vec{\theta}$ of a quantum gate sequence $U(\vec{\theta})$ are optimised for a given state $\rho$. Ideally, after optimisation, the sequence $U(\vec{\theta}_\text{opt})$ satisfies
	\begin{equation*}
		\rho' = U(\vec{\theta}_\text{opt})\rho U(\vec{\theta}_\text{opt})^\dagger = \rho_\text{diag},
	\end{equation*}
	where $\rho_\text{diag}$ is $\rho$ diagonalised in its eigenbasis and $\vec{\theta}_\text{opt}$ are the optimal angles. Classical gradient-based methods such as SPSA and Gradient Descent, or gradient-free optimisation~\cite{powell1994direct, powell2006fast} can be utilised in the training process. (2) In the eigenvalue readout subroutine, using the optimised unitary $U(\vec{\theta}_\text{opt})$ and one copy of state $\rho$, one can extract all the eigenvalues for low-rank states or the largest eigenvalues for full-rank states. This is achieved by measuring $\rho'$ in the computational basis, $\mathbf{b}=b_1b_2\ldots b_n$, as follows
    \begin{equation*}
        \lambda' = \bra{\mathbf{b}}\rho'\ket{\mathbf{b}},
    \end{equation*}
    where $\lambda'$ are inferred eigenvalues. (3) In the final step, we prepare the eigenvectors associated with the largest eigenvalues. If $\mathbf{b}'$ is a bit string associated with $\lambda'$, then the inferred eigenvectors $|v_{\mathbf{b}'}'\rangle$ are obtained as follows
    \begin{equation*}
        |{v}'_{{\mathbf{b}}'}\rangle = U(\theta_\textrm{opt})^\dagger\ket{{\mathbf{b}}'} = U(\theta_\textrm{opt})^\dagger\left(X^{b_1}\otimes\ldots\otimes X^{b_n} \right)\ket{\mathbf{0}}.
    \end{equation*}

In the reinforcement learning enhanced variational quantum state diagonalisation, an RL-agent is utilised to find the optimal configuration of $U(\vec{\theta})$. The agent contains a double deep-Q network (DDQN)~\cite{van2016deep} with $\epsilon$-greedy policy. 
The DDQN optimises the stability of Q-learning by using two networks to decouple the action selection $a'$ in the subsequent state $s'$ from its current reward $r$ evaluation, leading to less overestimated Q-values and more stable and reliable learning outcomes.

The key update equation is 
\begin{multline}
Q(s, a; \theta) \leftarrow Q(s, a; \theta) \\
+ \alpha [r + \gamma Q(s', \arg\max_{a'}Q(s', a'; \theta); \theta') - Q(s, a; \theta)]
\end{multline}

where, $Q(s, a; \theta)$ is the predicted Q-value for state $s$ and action $a$, parameterised by weights $\theta$ of the online network.
$\alpha$ is the learning rate and $\gamma$ is the discount factor. The settings of the parameters are similar to the ref.~\cite{kundu2024enhancing}.

The optimal policy is obtained by optimising the weights of the DDQN using the ADAM optimiser. In the RL setup, the RL-state encodes the ansatz and is defined by a tensor-based encoding presented introduced in~\cite{patel2024curriculum}. After each environment and agent interaction, the reward function
\begin{equation}
    R = \left\{\begin{array}{ll}
        +\mathcal{R} & \text{for } C_t(\vec{\theta})<\zeta+10^{-5}\\
        -\textrm{log}\left(C_t(\vec{\theta})-\zeta\right) & \text{for } C_t(\vec{\theta})>\zeta
        \end{array}\right.,\label{eq:log_reward}
\end{equation}
is calculated. The goal of the RL-agent is to reach the minimum error for a predefined threshold $\zeta$ (which is a hyperparameter of the model) by optimising the parameters of the $U(\vec{\theta})$ using the COBYLA optimiser. The cost function 
\begin{equation}
C_t(\vec{\theta}) = \text{Tr}(\rho^2) - \text{Tr}(\rho_t(\vec{\theta})^2).
\end{equation}
at each step $t$ is calculated for the ansatz which outputs a state $\rho_t(\vec{\theta})$.

\section{Analysis of the Haar random input quantum states}~\label{appndix:input-state-analysis}
In this section, we analyse the 2-qubit Haar random quantum states. \textcolor{black}{The Haar random states can be efficiently simulated using polynomial-time algorithms, as demonstrated in prior research~\cite{alagic2020efficient}. This holds even in scenarios where adversaries have unbounded capabilities with black-box oracle access~\cite{chen2024power}.} Here, we sample $100000$ Haar random quantum states and observe how the eigenvalues and the qubits' conditional entropy change with the state's entanglement.
    \begin{figure}[H]
    \centering
    \includegraphics[width=\linewidth]{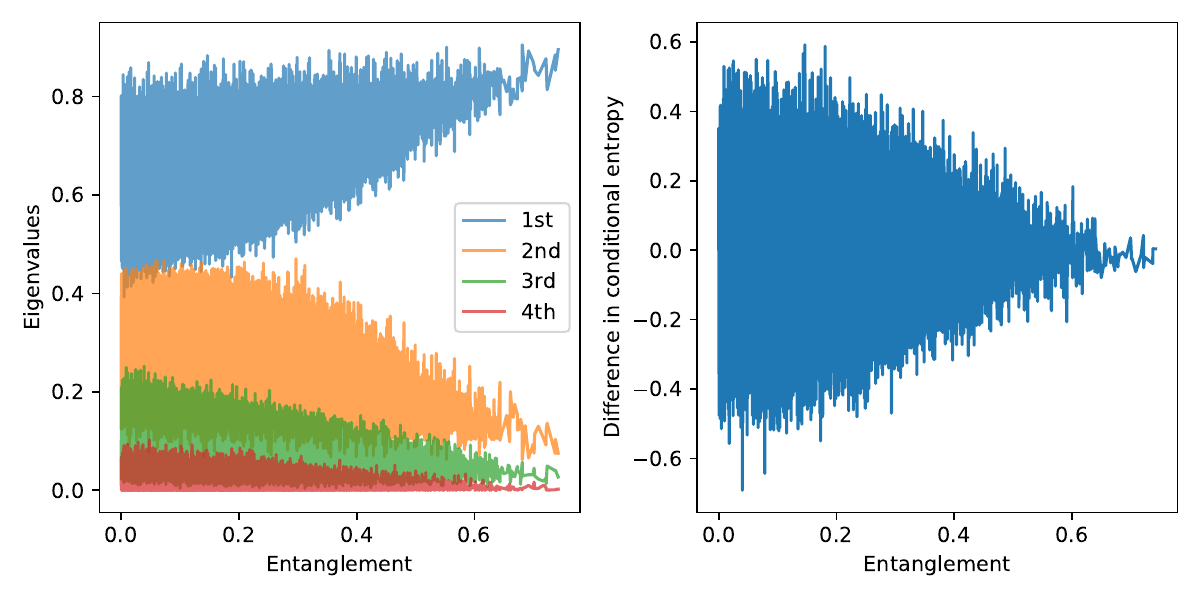}
    \caption{The difference in conditional entropy converges towards the zero as the entanglement of the sampled input state increases. Meanwhile, as the input state gets closer to the maximally entangled state (with concurrence $1$), the magnitude of the largest eigenvalue increases while the other eigenvalue decreases.}
    \label{fig:sampled-state-analysis}
    \end{figure}

In Fig.~\ref{fig:sampled-state-analysis}, see that the difference in conditional entropy in qubits of the input state varies evenly on either side of $S_{AB}^{\rho_{AB}}=0$ where $A$ and $B$ are subsystems containing one qubit each and $\rho_{AB}$ is the sampled Haar random state. Meanwhile, the area of the distribution of $S_{AB}^{\rho_{AB}}$ on \textcolor{black}{either} side of zero entropy converges towards zero as the input state entanglement increases. Furthermore, the magnitude of the input quantum state's largest eigenvalue increases with the increase in entanglement, and all the other eigenvalues converge towards $0$. It also shows that even for $100000$ seeds, the probability of sampling states with concurrence greater than $0.6$ is very low, justifying the low density of points after $0.6$ in Fig.~\ref{fig:anti_correlation_ent_bound}.

\section{Main results} \label{sec:mainResults}
In this section, we investigate the upper and lower bounds of concurrence of the ansatzes produced by the RL agent in diagonalising $2$-qubit random quantum states sampled from \texttt{IBM qiskit}'s \texttt{random\_density\_matrix} module. For the simulation, we consider multiple such quantum states (specified in the caption of figures), and for each state, \textcolor{black}{we use $10000$ episodes of RL-VQSD algorithm} to diagonalise the state. 
 \textcolor{black}{The algorithm consists of  DDQN with $\epsilon$-greedy policy where the $\epsilon$ decays in each step by a factor of $0.99995$ from its initial value $1$, down to a
minimal value $0.05$. Furthermore, the network is trained by ADAM optimizer~\cite{kingma2014adam} with a learning rate of 0.0001. The neural network consists of 5 hidden layers with 1000 neurons and a buffer size of 15000.}
The configuration of the RL-VQSD is provided in detail in Section~\ref{appndx:rl-vqsd}.
\subsection{Entanglement} \label{sec:entanglement}

\subsubsection{DDQN weight independent entanglement bound} \label{sec:DQNentBound}

To benchmark the performance of the RL-agent for a specific problem, it is ideal to take multiple initialisations of the DDQN (see Section~\ref{appndx:rl-vqsd} for a brief description). 
\begin{table}[H]
    \centering
    \begin{tabular}{c|c  c|c c}
       state & \multicolumn{2}{c|}{max. concurrence} & \multicolumn{2}{c}{min. concurrence} \\
       no. & avg. & std. div & avg. & std. div. \\
       \hline
       1 & 0.874 & 0.002 & 0.183 & 0.004 \\
       2 & 0.849 & 0.002 & 0.108 & 0.028 \\
       3 & 0.975 & 0.001 & 0.084 & 0.006 \\
       4 & 0.936 & 0.002 & 0.118 & 0.009 \\
       5 & 0.880 & 0.001 & 0.188 & 0.013 \\
       6 & 0.911 & 0.003 & 0.156 & 0.024 \\
       7 & 0.933 & 0.001 & 0.235 & 0.013 \\
       8 & 0.928 & 0.005 & 0.208 & 0.013 \\
       9 & 0.831 & 0.011 & 0.302 & 0.010 
    \end{tabular}
    \caption{\small The average and standard deviation of the final state's maximum (max. concurrence) and minimum concurrence (min. concurrence) after evolution through the circuit for five different initialisations of the DDQN weights for different quantum seed states. To obtain the results, we run RL-VQSD to diagonalize $2$-qubit random quantum states for $10000$ episodes. The agent then proposes $E_s<10000$ admissible ansatzes, which generate $E_s$ states after the random quantum state evolves them. 
    The max. (min.) concurrence is attributed to the ansatz that generates a state with the highest (lowest) concurrence following the evolution of the random quantum state.}
    \label{tab:DQNIndep}
\end{table}
To efficiently investigate the upper and lower bounds of concurrence~\cite{HW97,Wooters98} of the admissible ansatz, we calculate the \textit{concurrence of \textcolor{black}{the} admissible ansatz}, which is defined by the entanglement induced by the admissible ansatz when initialised with $\ket{00}$.
We first sample different random quantum states, then diagonalise them using RL-VQSD~\cite{kundu2024enhancing} for five different initialisations of DDQN weights. Thereafter, the agent suggests different possible configurations for the RL-ansatz, which diagonalises the sampled random quantum states. For a specific state, we collect such valid quantum circuits and find the concurrence of the state obtained after evolving through the RL-ansatz. It should be noted that finding the concurrence of the evolved state indicates how much the RL-ansatz induces entanglement to the input sampled state. See Section~\ref{appndix:input-state-analysis} for further analysis of the sampled random quantum states, where we show that in \texttt{qiskit}, \texttt{random\_density\_matrix} even for $100000$ seeds, the probability of sampling states with concurrence greater than $0.6$ is very low.

The results are presented in Table~\ref{tab:DQNIndep}. 
We observe that the optimal configuration of the RL-ansatz always gives us a state with concurrence close to one. Hence, the deviation in the maximum concurrence over different weights of the DDQN is more significant than the deviation in the minima of the concurrence. As seen in Table~\ref{tab:DQNIndep}, the deviation in the max concurrence is in the order of $10^{-3}$.
The low standard deviation implies that there is a negligible deviation in concurrence across different initialisation of the DDQN for a specific state. This indicates that the bounds on concurrence are invariant for different configurations of the RL-ansatz. Hence, proving that \textit{the entanglement bounds of the RL-ansatz are independent of initialisation of DDQNs}. Due to this observation, it is sufficient to consider a single initialisation of DDQN to benchmark the outcomes in the remaining simulations.

In the following subsection, we investigate how the upper and lower bounds on the entanglement of the RL-ansatz depend on the entanglement of the input quantum state.

\subsubsection{Anti-correlation between entanglement bounds} \label{sec:correlationEntBound}

Corresponding to a predefined threshold, the agent proposes different architectures of quantum circuits that diagonalise the quantum state. To observe the amount of entanglement generated by the RL-ansatz we calculate the concurrence of the state after passing through the RL-ansatz, just before the dephasing operation. By sorting out the quantum circuit with maximum and minimum entanglement, we observe their variation with the entanglement of the input state. It has been proved in the previous section that the upper and lower bounds of entanglement are independent of different initialisation of DDQNs; hence, for further benchmarking, we can constrain ourselves to just one initialisation of the DDQN weights.

\begin{figure}[H]
    \includegraphics[scale=0.47]{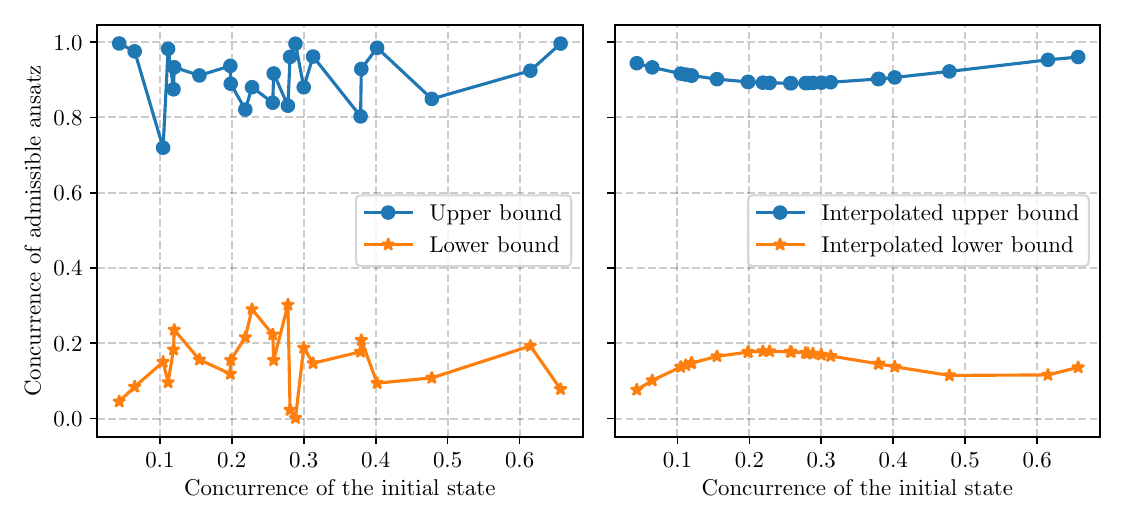}
    \caption{\small Reciprocal behaviour in the upper and lower bounds of concurrence with respect to the increasing concurrence in the input random state. For the first eleven points, the Pearson correlation coefficient (PCC) is $-0.982$, denoting a strong anti-correlation between the upper and lower bounds of concurrence. The strong anti-correlation turns into a mild correlation after concurrence of the initial state surpasses $0.322$. The correlation after the first eleven points changes from $-0.055$ to $0.049$. There is a phase transition from weak anti-correlation to weak correlation.}
    \label{fig:anti_correlation_ent_bound}
\end{figure}

In {Fig.~\ref{fig:anti_correlation_ent_bound}}, we illustrate the variation of the entanglement bounds with the entanglement of the input quantum state. To investigate the dependency, we use the Pearson correlation coefficient (PCC)~\cite{RN88} as a quantifier of correlation. See Section~\ref{app:preliminary} for a brief description. 
We see that the upper and lower bounds of entanglement are initially strongly anti-correlation when the input state entanglement is below $0.322$ and then strongly correlate afterwards, indicating a \textit{phase transition} in the upper and lower bounds of entanglement.

\begin{figure}[H]
    \centering
    \includegraphics[scale=0.5]{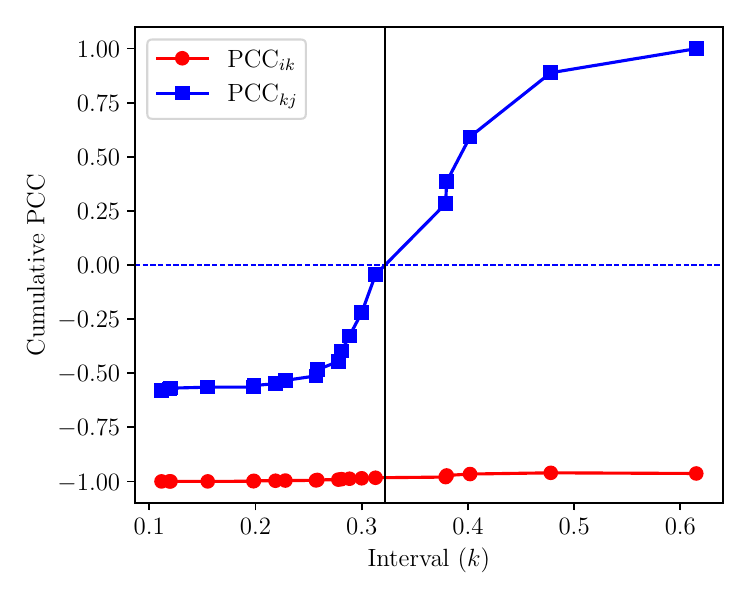}
    \caption{\small We present the phase transition of correlation between the upper and lower bounds of concurrence for the admissible RL-VQSD ansatz. To show this, we split the total range of the concurrence of the input state in two intervals $\mathbbm{u}_1 \coloneqq [i,k)$ and $\mathbbm{u}_2 \coloneqq [k,j]$, where $k$ is a free parameter and plot the variation of the Pearson correlation coefficient between the upper and lower bound on the concurrence of the ansatz corresponding to these two intervals. We denote by $\textrm{PCC}_{ik}$ and $\textrm{PCC}_{kj}$ the correlation coefficients in the intervals $\mathbbm{u}_1$ and $\mathbbm{u}_2$ respectively. The phase transition point is observed at $k=0.322$ where the concurrence bounds change from mild anti-correlation to mild correlation for the interval $\mathbbm{u}_2$ while remaining strongly anti-correlated in the interval $\mathbbm{u}_1$.} 
    \label{fig:anti_correlation_ent_bound_1}
\end{figure}

\textcolor{black}{Intuitively, when the entanglement of the starting state is low i.e. $k\leq0.322$, it becomes easier for the dephasing subroutine of the VQSD algorithm to eliminate the off-diagonal terms. Hence the agent proposes \textcolor{black}{an} admissible ansatz to either convert the state into a separable state (eliminating the need for a dephasing gate at the end) or increase its entanglement to the maximum level, thereby making full use of the dephasing gate's potential. This dynamic explains the observed anti-correlation between the entanglement bounds.}

\textcolor{black}{As noted previously, when the entanglement of the starting state is high (in the region $k\geq0.322$), it becomes easier for the dephasing subroutine to eliminate the off-diagonal terms. Thus for states with high entanglement, the agent proposes an admissible ansatz to maximize the entanglement of the starting state. Hence we observe the positive relation between the entanglement bounds.}


Next, we obtain the variation of PCC between the upper and lower bound on the concurrence of the initial state. Specifically, we introduce a parameter $k \in (i,j)$ to split the range of initial state concurrence $\mathbbm{u} \coloneqq [i,j]$ into two intervals: $\mathbbm{u}_1 \coloneqq [i,k)$ and $\mathbbm{u}_2 \coloneqq [k,j]$. Also, let us denote the set of lower and upper bounds on the concurrence of the ansatz as $u$ and $l$, respectively. 
For both the ranges $\mathbbm{u}_1$ and $\mathbbm{u}_2$, we introduce the following correlation function:
\begin{eqnarray}
    \eta^k_{w_{ik}} = w_{ik}\textrm{PCC}_{ik}+ (1-w_{ik})\textrm{PCC}_{kj},
    \label{eq:correlation_function}
\end{eqnarray}
where $w_{ik} \in \{0,1\}$ and $\textrm{PCC}_{ik}$, $\textrm{PCC}_{kj}$ are the value of PCC between the upper and lower bounds of concurrence of the ansatz for the intervals $\mathbbm{u}_1$ and $\mathbbm{u}_2$ respectively. For the sake of simplicity we denote $\textrm{PCC}_{ik}$ as $\textrm{PCC}_{ik}^{ul}$ and also $\textrm{PCC}_{kj}$ as $\textrm{PCC}_{kj}^{ul}$.

\begin{note}
    We observe that depending on the values of $w_{ik}$, $\eta^k_{w_{ik}}$ can take values either $\textrm{PCC}_{ik}$ or $\textrm{PCC}_{kj}$. For $\eta^k_{w_{ik}} = \textrm{PCC}_{ik}$, we obtain the amount of (anti-)correlation between the lower and upper bounds of concurrence of the ansatz for the interval $\mathbbm{u}_1$. While for $\eta^k_{w_{ik}} = \textrm{PCC}_{kj}$, we obtain the amount of (anti-)correlation for the interval $\mathbbm{u}_2$. 
    We leave proving $\eta^k_{w_{ik}}$ is a valid measure of correlation for arbitrary values of $w_{ik} \in (0,1)$ as an open problem.
\end{note}
We present in {Fig.~\ref{fig:anti_correlation_ent_bound_1}}, the variation of $\eta_0^k$ (in blue) and $\eta_1^k$ (in red) for different values of $k$.  We observe that $\eta_1^k \approx -1~\forall k \in (i,j)$. This indicates strong cumulative anti-correlation between the upper and lower bounds of concurrence. For $\eta_0^k$, we observe a gradual change of $\text{PCC}_{kj}$ from -0.582 to 1. We note that $\eta_0^k$ changes from $-0.0451$ to $0.279$ for $k \in [0.313,0.379]$ with $\eta_0^k = 0$ at $k = k_\ast = 0.322$. We call this point the \textit{phase transition point of correlation between upper and lower bounds of concurrence} for the VQSD ansatz. At values of $k < k_\ast$, there is anti-correlation between the upper and lower bounds of concurrence of the ansatz for both the intervals $\mathbbm{u}_1$ and $\mathbbm{u}_2$. While for values of $k > k_\ast$, there is (anti-)correlation between the upper and lower bounds of concurrence of the ansatz for both the intervals $(\mathbbm{u}_2)~\mathbbm{u}_1$.

\textcolor{black}{The investigation in Fig.~\ref{fig:anti_correlation_ent_bound} and Fig.~\ref{fig:anti_correlation_ent_bound_1} suggests that the structure of the quantum state in terms of its entanglement properties can significantly influence the effectiveness of certain ansatz strategies. Specifically, when the entanglement is low i.e. below $0.322$, the variability in the entanglement bounds can be exploited. For wide entanglement bounds, the flexibility of the ansatz allows for the generation of states that are either highly entangled or nearly separable, starting from $|0\rangle^{\otimes 2}$.}

\subsubsection{Enhancing the performance of agent based on initial state entanglement} \label{sec:agentEnvInteract}

Here, we provide a detailed investigation of the admissible ansatzes produced by the RL-agent in terms of ansatz depth, the total number of 2-qubit gates and the total number of gates in the upper and lower bounds of concurrence.

\begin{figure}[H]
    \centering
    \includegraphics[width=\linewidth]{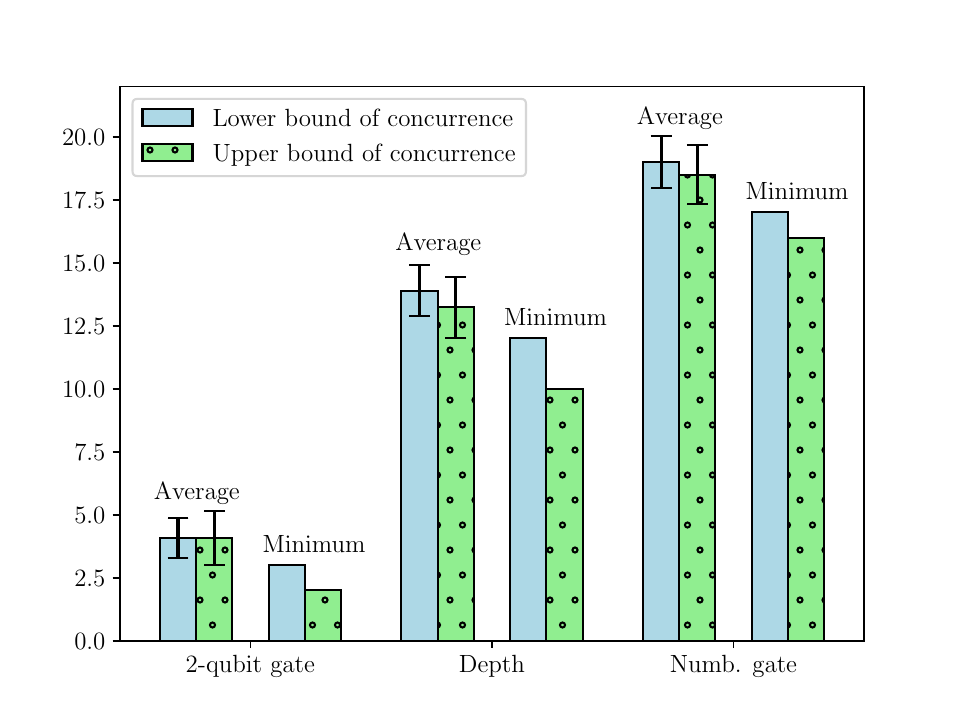}
    \caption{\textcolor{black}{\small The RL-agent can generate an admissible ansatz with strong entanglement with fewer 2-qubit gates (2-qubit gate) and circuit depth (Depth) and total number of gates (Numb. gate) compared to the ansatzes with weak entanglement. As the main aim of RL-VQSD is to diagonalise unitary with a small number of gates and depth, our analysis shows that if we particularly focus on the admissible ansatzes near the upper bound of concurrence, we can not only reduce the computational cost of RL-VQSD severely but \textcolor{black}{also} find \textcolor{black}{the} optimal admissible ansatzes}.}
    \label{fig:resource_on_bounds}
\end{figure}


Our investigation reveals that the {admissible ansatzes} corresponding to the upper bound of concurrence has, on average, a shorter depth of $13.23$ and a smaller number of gates of $18.5$ compared to the {admissible ansatzes} corresponding to the lower bound of concurrence where the depth is $13.9$ and number of gates is $19$ over $25$ random quantum states. Moreover, {Fig.~\ref{fig:resource_on_bounds}} makes it more prominent that the minimum number of gates and the depth of the admissible ansatzes are smaller for the upper bound of concurrence than the lower. This observation is crucial in reducing the subspace of all possible {admissible ansatzes}. The main motivation of RL-QAS in VQAs is to find ansatzes that solve a VQA with a minimum number of gates and circuit depth; in the case of RL-VQSD, this observation can help us to narrow down the search for an optimal admissible ansatz in the two following ways. 

\subsubsection*{a. Entanglement dependent initialisation of RL-state} \label{para:ent-dep-init}

\begin{figure}[H]
    \centering
    \includegraphics[scale=0.45]{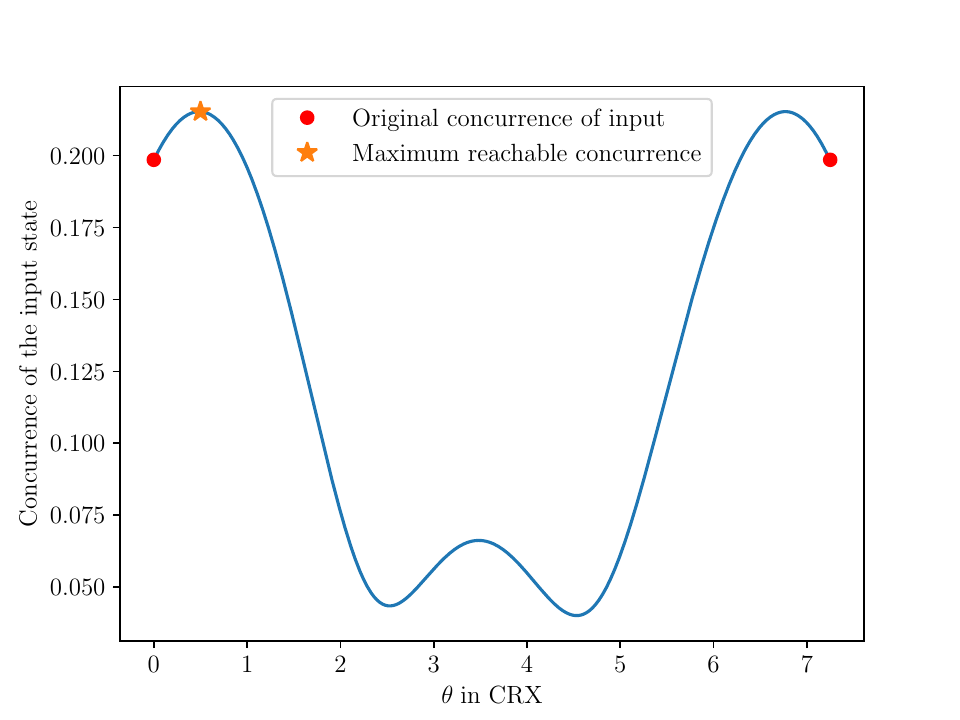}
    \caption{The concurrence of input state with $\theta$ of \texttt{CRX}($\theta$) gate in \textsc{Entanglement Enhancing} increases the concurrence of the input random state from $0.198$ to $0.215$ when $\theta=0.5$.}
    \label{fig:entanglement-enhancing-effect}
\end{figure}

In RL-VQSD~\cite{kundu2024enhancing}, the RL-state that encodes the ansatz initiates from an empty circuit (i.e., in circuit~\ref{circ:ent-dep-initialization} the \textsc{Entanglement Enhancing} block is absent and $U_{\text{RL-agent}}=\mathds{1}$). We showed that the optimal admissible ansatz, on average, lies in the upper bound of concurrence, motivating us to initialise the quantum state (that is diagonalised using RL-VQSD) closest to the upper bound of entanglement. To do this, the ansatz is divided into two parts 
\begin{equation}
\small
    \begin{quantikz} \lstick[2]{$\rho$} & \gate{H}\gategroup[2,steps=3,style={dashed,rounded corners, inner xsep=2pt},background,label style={label position=below,anchor=north,yshift=-0.2cm}]{{\sc\small Entanglement Enhancing Block}} & \ctrl{1} & \gate{H} & \gate[2]{U_{\text{RL-ansatz}}} & \\ 
    & & \gate{CRX(\theta)} & & & \end{quantikz},\label{circ:ent-dep-initialization}
\end{equation}
where the first, \textsc{Entanglement Enhancing} (EE) block, helps the diagonalising state to achieve the highest entanglement before it is fed to the second part, $U_{\text{RL-agent}}$, which the RL-agent decides. For example, for seed $27$ (see Section~\ref{appndix:input-state-analysis} for details), we get a random quantum state with concurrence $0.198$ (for seed $27$, and see Appendix~\ref{appndx:sampling_state} for details on how to generate Haar random state), and using the EE, we can increase the input state concurrence to $0.215$ at $\theta=0.5$ as shown in Fig.~\ref{fig:entanglement-enhancing-effect}.
Our investigation reveals that turning on the EE block with $\theta=0.5$ enhances the performance of the RL-agent by $2$-times compared to the default setting at $\theta=0$. 
The results are summarised in the Tab.~\ref{tab:agent-optimization} where we see that the number of successful episodes and the reward acquired (where the reward function is in the form defined in Eq.~\ref{eq:log_reward}) by the agent in the RL-VQSD process increase by $2$-fold while the EE block is activated for $\theta=0.5$ compared to default settings at $\theta=0$.
\begin{table}[H]
    \centering
    \caption{Turning on the \textsc{Entanglement Enhancing} (EE) mode (denoted as Enhanced ($\theta=0.5$)) enhances the learning of the performance of the RL-agent in the absence of the EE mode (denoted as Default ($\theta=0$)). In the first $3500$ episodes, the Enhanced ($\theta=0.5$) collects $2.08$ times more reward (tot. rwd. coll.) and $2.22$ times more successful episodes (Tot. succ. ep.) than Default ($\theta=0$). Moreover, the average number of 2-qubit (2q gate) and depth over the first $100$ successful episodes reduced in the case of Enhanced ($\theta=0.5$) settings compared to Default ($\theta=0$), showing enhanced learning of the RL-agent for entanglement enhanced input state.}
    \begin{tabular}{c|ccccc}
        EE mode & Tot. succ. & Tot. rwd. & \multicolumn{3}{c}{1st 100 succ. ep. (avg.)}\\
        ($\theta$) & ep. & coll. & 1q gate & 2q gate & depth\\
       \hline
        \makecell{Enhanced\\ ($\theta=0.5$)} & $858$ & $4.45\times10^{5}$ & $15.41$ & $4.02$ & $14.6$ \\
        \hline
        \makecell{Default\\ ($\theta=0$)} 
        & $387$ & $2.14\times10^{5}$ & $14.71$ & $5.29$ & $15.59$
    \end{tabular}
    \label{tab:agent-optimization}
\end{table}
Meanwhile, in the Tab.~\ref{tab:ansatz-optimization}, we show that the EE block also helps to optimise the cost function and the minimum number of 2-qubit gates in the optimal admissible ansatz compared to the default settings.
\begin{table}[H]
    \centering
    \caption{Turning on the \textsc{Entanglement Enhancing} (EE) mode (denoted as Enhanced ($\theta=0.5$)) optimises the number of minimum number of 2-qubit gates (Min. 2q gate) in RL-ansatz better than in the absence of the EE mode (denoted as Default ($\theta=0$)). Moreover, the Enhanced ($\theta=0.5$) mode minimises the cost function better than the Default ($\theta=0$) mode.}
    \begin{tabular}{c|cc c}
        EE mode ($\theta$) & \makecell{Cost\\ function} & \makecell{Min.\\ 1q gate} & \makecell{Min.\\ 2q gate} \\
       \hline
        Enhanced ($\theta=0.5$) & $5.12\times10^{-7}$ & $14$ & $2$\\
        Default ($\theta=0$) & $6.08\times10^{-7}$ & $11$ & $5$
    \end{tabular}
    \label{tab:ansatz-optimization}
\end{table}

It should be noted that due to the sake of the interpretability of the EE block, we chose the block resembling a maximally entangled circuit for a 2-qubit system. Where the entangler (i.e. the \texttt{CX}) gate is parameterised and can induce no entanglement (for $\theta=0$) and maximal entanglement (for $\theta=0.5$) to $|00\rangle$ state. The second Hadamard (\texttt{H}) gate on the first qubit is introduced to cancel out the first \texttt{H} in the absence of the entangler at $\theta=0$. We can automate the search for gateset in the EE block utilising an RL-agent and diagonalise the input state simultaneously by decomposing the RLQAS into two subroutines. Suppose the total number of steps in an episode is $S$, then in the first subroutine for $s_\text{EE}<S$ steps the agent's task would be to find a gateset (consists of maximum $s_\text{EE}$ gates) that maximises the entanglement of the input state by utilising a reward function $R_\text{EE}$. One can formulate $R_\text{EE}$ similar to ref.~\cite{kuo2021quantum}. 
Wherein in the second subroutine, for the remaining steps $S-s_\text{EE}$, the agent focuses on finding possible structures of $U_\text{RL-ansatz}$ to diagonalise the output state of the EE block using the reward defined in Eq.~\ref{eq:log_reward}.

The aforementioned observations lead us to the conclusion that maximising the entanglement of the input state through the initialisation strategy of Fig.~\ref{circ:ent-dep-initialization} enhances the learning process of the RL-agent and facilitates the further optimisation of the admissible ansatzes compared to the start-of-art~\cite{kundu2024reinforcement}.  

\subsection{Conditional quantum entropy} \label{sec:conditionalEntropy}

\subsubsection{Contribution of individual qubits for change in entanglement} \label{sec:entropyEntanglement}

We quantify the contribution of each qubit using the conditional quantum entropy $S_{q_0|q_1}^{\rho'_{q_0q_1}}$~\cite{CA97,CA99,wilde2013quantum}. The conditional entropy of an individual qubit for an $N$ qubit quantum system indicates the amount of uncertainty remaining about that qubit after measuring the whole system. Hence, it is essential to know how much information the state of qubits provides about the state of the $N$ qubits. For a detailed discussion and notation, see Section~\ref{app:preliminary}.

\begin{figure}[H]
\includegraphics[scale=0.5]{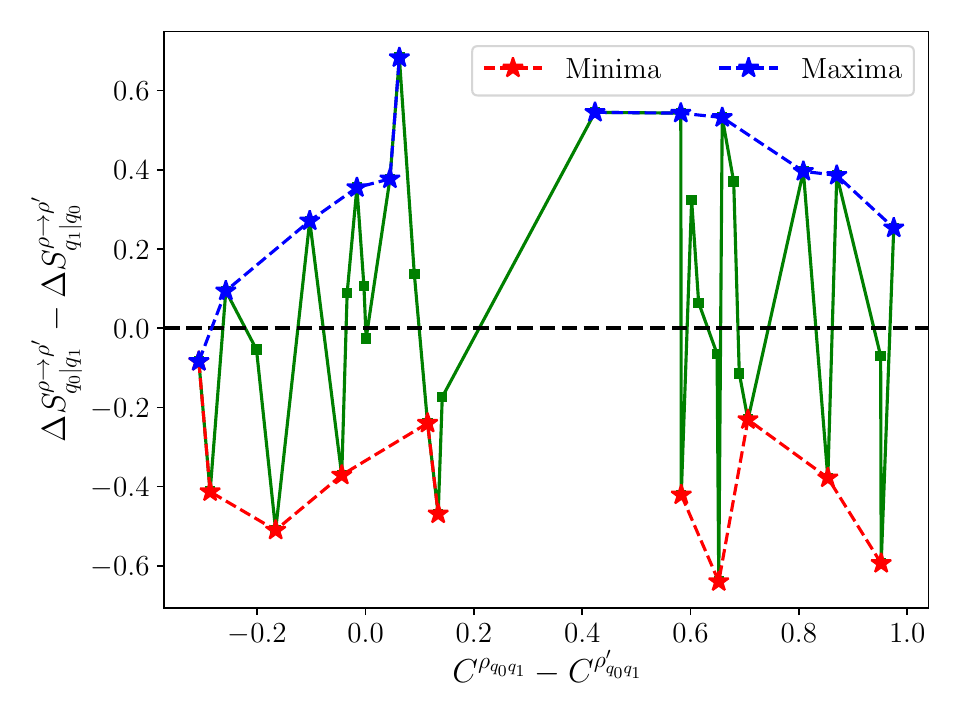}
\caption{\small \textcolor{black}{We analyze the relative change in the contribution of the first and second qubits as a function of changes in entanglement. To quantify the relative contribution of individual qubits to the RL-VQSD ansatz, we use $\Delta_{q_0}^{q_1}$ (see Eq.~\eqref{eq:contribution}). For the change in concurrence due to the ansatz, the distribution of $\Delta_{q_0}^{q_1}$ is uniform with respect to $\Delta_{q_0}^{q_1} = 0$. The cumulative weight of all data points in the region $\Delta_{q_0}^{q_1} < 0$ is 4.854, while that for data points in the region $\Delta_{q_0}^{q_1} > 0$ is 5.524. This indicates an equal contribution from both qubits to the RL-VQSD ansatz across the range of changes in entanglement}.} 
\label{fig:diffCondEntropyAnsatz}
\end{figure}
Let the change in the conditional entropy of the qubit $q_0 (q_1)$ after evolving through the VQSD ansatz be given by $\Delta S^{\rho \to{\rho'}}_{q_0|q_1} (\Delta S^{\rho \to{\rho'}}_{q_1|q_0})$. We quantify the relative contribution of individual qubits via the quantity:
\begin{eqnarray}
    \Delta_{q_0}^{q_1} \coloneqq \Delta S^{\rho \to{\rho'}}_{q_0 | q_1} - \Delta S^{\rho \to{\rho'}}_{q_1|q_0}. \label{eq:contribution}
\end{eqnarray}
We note that $\Delta_{q_0}^{q_1} > 0$ implies a greater change in conditional entropy of $q_0$ as compared to $q_1$, and hence the ansatz has a greater contribution from $q_0$ as compared to $q_1$. Following a similar reasoning $\Delta_{q_0}^{q_1} < 0$ implies $q_1$ has greater contribution as compared to $q_0$, while $\Delta_{q_0}^{q_1} = 0$ implies both $q_0$ and $q_1$ contributes equally to the ansatz. 

We present the variation of $\Delta_{q_0}^{q_1}$ for the change in concurrence due to the ansatz given by $\Delta_c \coloneqq C^{\rho_{q_0q_1}} - C^{\rho'_{q_0q_1}}$ in {Fig.~\ref{fig:diffCondEntropyAnsatz}}. We observe that the distribution of $\Delta_{q_0}^{q_1}$ is bounded in the interval $[-0.7,0.7]$ and is uniform with respect to $\Delta_{q_0}^{q_1} = 0$. The magnitude of the cumulative weight of all data points in the region $\Delta_{q_0}^{q_1} < 0$ is $4.854$ while that for data points in the region $\Delta_{q_0}^{q_1} > 0$ is $5.524$. This indicates equal contribution from both the qubits to the RL-VQSD ansatz for the range of change in entanglement. In the region of $\Delta_{q_0}^{q_1} > 0$, the relative contribution of $q_0$ first increases and then decreases. However, no such pattern is observed for the region of $\Delta_{q_0}^{q_1} < 0$. When $\Delta_c > 0.4$, the value of $\Delta_{q_0}^{q_1}$ monotonically decreases with increase in $\Delta_c$. Also, for $\Delta_c < 0.2$, $\Delta_{q_0}^{q_1}$ increases monotonically with increase in $\Delta_c$.

\subsubsection{Problem with variational quantum state diagonalisation} \label{sec:problem}

In this section, we investigate the contribution of individual qubits to the performance of the RL-VQSD algorithm. Through this investigation, we get a deeper insight into why the variational quantum state diagonalization algorithm can not find the smallest eigenvalues.

\begin{figure}[H]
    \centering
    \includegraphics[scale=0.7]{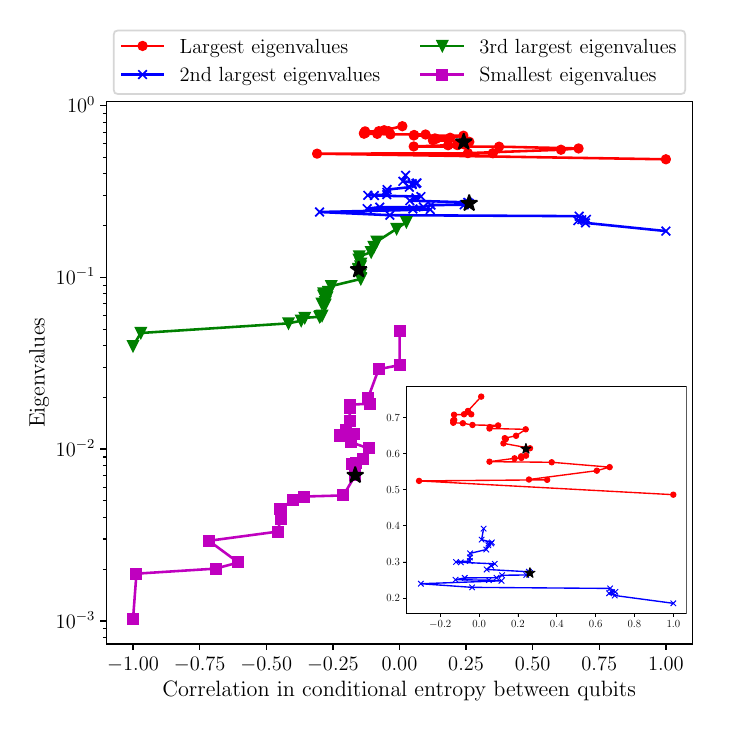}
    \caption{\small \textcolor{black}{On average over $35$ random quantum states the correlation in conditional entropy between the first and second qubits decreases as the magnitude of the largest and second largest eigenvalues increases. Meanwhile, the same correlation between the qubits increases with the magnitude of the third largest and the smallest eigenvalues. The RL-ansatz proposes specific structures of admissible ansatzes, and for each such ansatz, it is not possible to have both types of correlations simultaneously. Hence, our investigation suggests that it is not feasible to find the smallest eigenvalues using the same ansatz that provides a good approximation of the largest eigenvalues. The correlation of conditional entropy between qubits is calculated using the Pearson correlation coefficient (PCC) and is the median over all configurations of RL-ansatz proposed by the agent. The $\bigstar$ signifies the eigenvalues corresponding to the same state.}}
    \label{fig:cond_ent_pear_corr_low_ent}
\end{figure}

We present the main observations in {Fig.~\ref{fig:cond_ent_pear_corr_low_ent}}, where we plot the four eigenvalues of $35$, $2$-qubit Haar random mixed quantum states with the correlation in conditional entropy among the qubits. For an individual quantum state, the RL-agent proposes $\mathbf{s}$ different structures of \textcolor{black}{the} {admissible ansatzes}. Then we get $\rho_i^\prime$ states where $i\in [0,\mathbf{s}]$. Afterwards, we calculate the conditional entropy of individual qubits for each state and take the median.\footnote{Calculating the median over the mean is justified by the broad range of conditional entropy distribution, which makes the median a more appropriate measure of central tendency.} Then, we calculate the correction in conditional entropy among the qubits using PCC.

In {Fig.~\ref{fig:cond_ent_pear_corr_low_ent}}, we observe that a mild anti-correlation between the qubits is required to find the two largest eigenvalues. As the magnitude of the largest eigenvalues decreases, the mild anti-correlation turns into a strong correlation between the qubits. \textcolor{black}{This indicates that the qubits require mild to strong correlation to achieve the two largest eigenvalues}.
Meanwhile, \textcolor{black}{we require a mild to strong anti-correlation between the qubits to find the smallest eigenvalues of the same states}. Then, as the magnitude of the smallest eigenvalue increases, the strong anti-correlation turns into mild anti-correlation. 

We recall that the correlation is calculated among the qubits based on the conditional entropy of individual qubits. As it is impossible to have two different kinds of correlation \textcolor{black}{(such as mild correlation and mild anti-correlation simultaneously)} between the qubits in the same ansatz structure, we can not find the largest and smallest eigenvalues utilising the same ansatz. For example, we mark the eigenvalues of a state with $\bigstar$, where we observe that to find the largest two eigenvalues, the RL-ansatz need to induce a \textcolor{black}{mild correlation} between the two qubits; meanwhile, to obtain the smallest eigenvalues, we require \textcolor{black}{mild anti-correlation}.

\begin{figure}[H]
    \centering
\includegraphics[width=\linewidth]{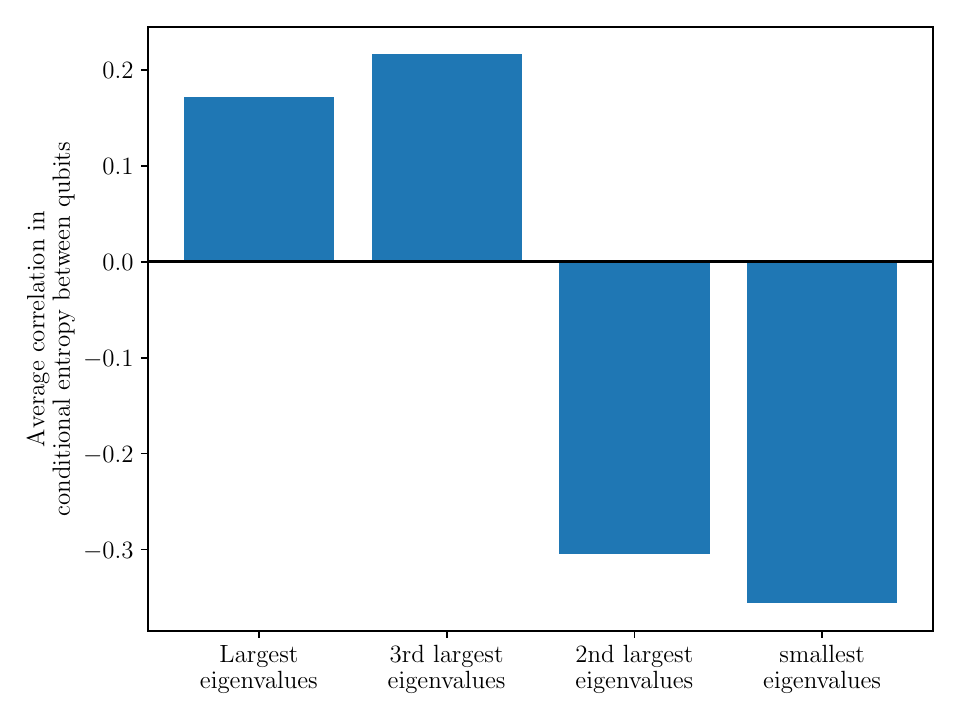}
    \caption{\textcolor{black}{\small The average correlation in conditional entropy among qubits is in the range of mild correlation for $35$ random quantum states for the first two largest eigenvalues whereas the conditional entropy among qubits requires a mild anti-correlation for the smallest eigenvalues. This observation, along with the observation in {Fig.~\ref{fig:cond_ent_pear_corr_low_ent}} certify that with a specific configuration of \textcolor{black}{an admissible ansatz}, we can not simultaneously find the largest and the smallest eigenvalues.}}
    \label{fig:average_anti_correlation}
\end{figure}
\textcolor{black}{The average correlation in conditional entropy among qubits falls within the range of mild correlation for the first two largest eigenvalues across $35$ random quantum states. Conversely, the conditional entropy among qubits exhibits a mild anti-correlation for the smallest eigenvalues. This finding, together with the observation in {Fig.~\ref{fig:average_anti_correlation}}, indicates that with a specific configuration of an admissible ansatz, it is interactable to simultaneously identify the largest and smallest eigenvalues. It should be noted that for the $35$ random quantum states, the average correlation in conditional entropy lies in the mild correlation zone, indicating that RL-VQSD can efficiently find the largest few eigenvalues. However, it fails to provide a good approximation for the smallest eigenvalues simultaneously.}


\section{Discussion} \label{sec:discussion}
In this article, we use tools from quantum information theory to analyse the {admissible ansatzes} proposed by the RL-agent for solving reinforcement learning-based quantum architecture search problems, and we particularly investigate the RL-agent proposed ansatzes in solving recently proposed RL-VQSD problem. 

We observe that the concurrence of the admissible ansatz ranges between an upper and lower bound, which are independent of the initialisation of the weights of the deep-Q network. The upper and lower bounds are initially anti-correlated with respect to the initial state entanglement. The anti-correlation turns into a mild and eventually strong correlation as the entanglement of the initial state surpasses beyond a phase transition point at concurrence $0.322$. We also see that the optimal configuration of the admissible ansatz lies in the upper bound of concurrence, which has a smaller circuit depth and requires fewer gates than the lower bound. These observations provide insight regarding the relation between the entanglement of the initial state and the entanglement generated by the RL-VQSD admissible ansatzes. We effectively utilise these observations to greatly reduce the RL-agent and environment interaction, i.e., the computational time needed to solve the RL-VQSD problem. 
Specifically, we propose an entanglement enhancing block that adjusts the input state's concurrence thereby enhancing the performance of the RL-agent by a factor of $2$.
Additionally, the admissible ansatzes not only diagonalise the random quantum states with high accuracy but can be used to generate close to maximum entangled states from the vacuum state. 

Furthermore, we quantify the contribution of each qubit in the admissible ansatz using conditional quantum entropy. For the admissible RL-VQSD ansatz, we observe an equal contribution from each qubit for the full range of change in the entanglement of the state due to the ansatz. 
Focusing on the task of obtaining the eigenvalues of the starting state, we observe that a \textcolor{black}{mild correlation} between the qubits of the admissible ansatz is required to obtain the two largest eigenvalues, while \textcolor{black}{mild anti-correlation} between the qubits is needed to find the smallest eigenvalues of the same state. Noting that it is impossible to have both mild correlation and anti-correlation between the qubits of the same ansatz, we cannot simultaneously obtain both the largest and smallest eigenvalues from the RL-VQSD admissible ansatz. From the observation, most of the admissible ansatz has \textcolor{black}{mild correlation} among the qubits, which explains why it is easier for the VQSD algorithm to find the largest eigenvalues than the smallest ones. 

\textcolor{black}{For future work, one could conduct a comprehensive study of the implication of introducing correlations such as entanglement~\cite{HHHH09}, nonlocality~\cite{brunner2014bell,sadhu2023testing} or steering~\cite{uola2020quantum,sadhu2024steering}-based reward functions of the RL-agent in QAS.} Specifically, given that most of the admissible ansatzes have high values of concurrence, it is expected that an entanglement or non-locality-based reward function may further reduce the time required to find the optimal configurations. Also, we may expect this to provide an optimal quantum circuit that provides both the eigenvalues of a random quantum state with higher accuracy within a smaller time. 

\textcolor{black}{In this work, we present an intuitive explanation of the phase transition behavior of the bounds on the entanglement of the admissible ansatz. However, we point out that the phase transition point is problem-specific. It would be interesting to see if the phase transition behavior still holds for other classes of algorithms, a problem we leave for future research.}

\textcolor{black}{Furthermore, using a similar approach, presented in our paper one can rigorously investigate the RL-ansatz in the task of minimizing the entanglement entropy of a quantum many-body system to identify quantum phase transition~\cite{an2021learning} and finding the ground state of chemical hamiltonian~\cite{patel2024curriculum,ostaszewski2021reinforcement}}.

\begin{acknowledgements}
A. Sadhu acknowledges the Ph.D. fellowship from the Raman Research Institute, Bangalore. AK would like to thank Jaros\l{}aw Miszczak and Mateusz Ostaszewski for their valuable comments on the manuscript and discussions. This research was partially supported by PL-Grid Infrastructure Grant Number PLG/2023/016205.
\end{acknowledgements}

\appendix

\section{Preliminaries} \label{app:preliminary}
We assume that every quantum system say $S$ has an associated Hilbert space $\mathscr{H}_S$ of finite dimension $d_S$. The state of a quantum system $S$ is represented by a density operator $\rho_S$ defined on $\mathscr{H}_S$ and satisfies the conditions (a) Positivity: $\rho_S \geq 0$, (b) Hermitian: $\rho_S = \rho_S^\dagger$ and (c) Trace Preserving: $\Tr[\rho_S] = 1$. We denote the set of density operators of $S$ as $\mathscr{D}(\mathscr{H}_S)$. A pure state is a rank-one density operator given by $\rho_S \coloneqq \op{\psi}_A$, where $\ket{\psi}_S \in \mathscr{H}_S$ is a unit-norm vector in $\mathscr{H}_S$. We denote the density operator of a composite system $S_1S_2$ as $\rho_{S_1S_2} \in \mathscr{H}_{S_1S_2}$ with $\Tr_{S_2} \rho_{S_1S_2} = \rho_{S_1} \in \mathscr{H}_{S_1}$ being the reduced state of $S_1$. The state $\rho_{S_1S_2} \in \mathscr{H}_{S_1S_2}$ is called separable if it cannot be expressed in the form of $\rho_{S_1S_2} = \sum_x p_x \rho_{S_1}^x \otimes \rho_{S_2}^x$, where $\{\rho_{S_1}^x\}_x$ and $\{\rho_{S_2}^x\}_x$ are sets of pure states, $p_x \in [0,1]$ and $\sum_x p_x = 1$. The states that cannot be expressed in the above form are said to be entangled~\cite{HHHH09}. Next, we briefly discuss a relevant measure of entanglement for this work. 

\subsection{Concurrence} \label{app:conc}
We quantify the amount of entanglement present in a bipartite quantum state $\rho_{AB} \in \mathscr{D}(\mathscr{H}_A \otimes \mathscr{H}_B)$ using concurrence~\cite{HW97,Wooters98} which is defined as:
\begin{eqnarray}
    C^{\rho_{AB}} = \max \{0,\lambda^e_1-\lambda^e_2-\lambda^e_3-\lambda^e_4\}
\end{eqnarray}
where $\lambda^e_s$ are the square root of the eigenvalues of $\rho_{AB} \Tilde{\rho}_{AB}$ in descending order with $\Tilde{\rho}_{AB} = (\sigma_y \otimes \sigma_y) \rho^*_{AB} (\sigma_y \otimes \sigma_y)$ being the spin-flipped state of $\rho_{AB}$ and $\sigma_y$ being a Pauli spin matrix. For the state $\rho_{AB} = \op{\psi}_{AB}$, the concurrence is defined as
\begin{eqnarray}
    \mathcal{C}^{\ket{\psi}_{AB}} = \sqrt{2 [1 - \Tr \rho_A^2]},
\end{eqnarray}
where $\rho_A = \Tr_B \rho_{AB}$ is a subsystem of the combined state. It is easy to see that the concurrence of a pure bipartite qubit maximally entangled state is one, while that of a bipartite qubit separable state is zero. In the following subsection, we discuss a measure to quantify the contribution of individual qubits in a quantum circuit. 

\subsection{Conditional quantum entropy}\label{app:CondQEnt}

We quantify the contribution of each qubit in a quantum circuit using the conditional quantum entropy~\cite{CA97,CA99}, which is defined as follows. 
\begin{definition} [Conditional quantum entropy~\cite{wilde2013quantum}]
    For the state $\rho'_{AB} \in \mathscr{D}(\mathscr{H}_A \otimes \mathscr{H}_B)$, the conditional quantum entropy $S_{A|B}^{\rho'_{AB}}$ is equal to the difference of the joint quantum entropy $S^{\rho'_{AB}}_{AB}$ and the marginal entropy $S^{\rho'_{AB}}_B$:
    \begin{equation}
        S_{A|B}^{\rho'_{AB}} \coloneqq S^{\rho'_{AB}}_{AB} -S^{\rho'_{AB}}_B.
    \end{equation}
\end{definition}
Operationally, the conditional entropy $S_{A|B}^{\rho'_{AB}}$ of the state for the state $\rho'_{AB}$ signifies how mixed the marginal state $\rho'_{A}$ ($\rho'_{B}$) is as compared $\rho'_{AB}$. For more details of the following theorems and observations, please refer~\cite{wilde2013quantum}.
\begin{observation}
    For the state $\rho'_{AB} \in \mathscr{D}(\mathscr{H}_A \otimes \mathscr{H}_B)$, consider a purification $\ket{\psi}_{ABE} \in  \mathscr{H}_A \otimes \mathscr{H}_B \otimes \mathscr{H}_E$. It then follows that:
    \begin{eqnarray}
        S_{A|B}^{\rho'_{AB}} = S_E^\psi - S_B^\psi.
    \end{eqnarray}
    The conditional quantum information thus measures the difference in the entropy of the states $\rho_E \in \mathscr{D}(\mathscr{H}_E)$ and $\rho_B \in \mathscr{D}(\mathscr{H}_B)$.
\end{observation}

In the following subsection, we discuss a measure to quantify the amount of correlation between two random variables. 

\subsection{Correlation coefficient}

We quantify the amount of correlation between two random variables $r_1$ and $r_2$ using the Pearson correlation coefficient ($\textrm{PCC}$)~\cite{RN88} defined as:
\begin{eqnarray}
    \textrm{PCC}^{r_1r_2} = \frac{\mathrm{E}[(r_1 - \mu_{r_1})(r_2 - \mu_{r_2})]}{\sigma_{r_1} \sigma_{r_2}},
\end{eqnarray}
where $\mu_{r_j}$, $\sigma_{r_j}$ are the mean and standard deviation of the distribution $r_j$ with $j \in \{1,2\}$ and $\mathrm{E}[.]$ is an operator that returns the expectation value of a random variable. Note that the Pearson correlation coefficient (PCC) is symmetric: $\textrm{PCC}^{r_1r_2} = \textrm{PCC}^{r_2r_1}$. The value of $\textrm{PCC}$ ranges in between $-1$ to $+1$. A correlation value of $+1 (-1)$ implies perfect correlation (anti-correlation) between the two random variables $r_1$ and $r_2$, while zero implies no linear dependence between the variables.

\subsection{Sampling a random quantum state}
\label{appndx:sampling_state}

If $n$ is the number of qubits and $s$ is the random seed to sample, then we can sample a state from \texttt{random\_density\_matrix} of \texttt{IBM qiskit}'s \texttt{quantum\_info} module as follows
\begin{equation}
\texttt{random\_density\_matrix(2**n, seed=s)}.
\end{equation}
The state used in~\ref{para:ent-dep-init}, is sampled using $n=2$ and $s=27$.

\bibliography{ref}

\end{document}